\documentclass[prb,aps,twocolumn,showpacs]{revtex4}
\usepackage{dcolumn}
\usepackage{amsmath}
\usepackage{graphicx}

\begin{document}

    %draft

\title{ Measurement of the spin polarization of the magnetic semiconductor EuS with zero-field and Zeeman-split Andreev reflection spectroscopy}

\author{Cong Ren$^{*}$, J. Trbovic, R. L. Kallaher, J. G. Braden, J.S. Parker, S. von Moln\'{a}r, and P. Xiong$^{\#}$}

\affiliation{Department of Physics and Center for Materials Research
and Technology (MARTECH), Florida State University, Tallahassee, FL
32306}

\begin{abstract}
We report measurements of the spin polarization
(\textbf{\textit{P}}) of the concentrated magnetic semiconductor EuS
using both zero-field and Zeeman-split Andreev reflection
spectroscopy (ARS) with EuS/Al planar junctions. The zero-field ARS
spectra are well described by the modified (spin-polarized) BTK
model with expected superconducting energy gap and actual
measurement temperature (no additional spectral broadening). The
fittings consistently yield \textbf{\textit{P}} close to 80\%
regardless of the barrier strength. Moreover, we performed ARS in
the presence of a Zeeman-splitting of the quasiparticle density of
states in Al. To describe the Zeeman-split ARS spectra, we develop a
theoretical model which incorporates the solution to the Maki-Fulde
equations into the modified BTK analysis. The method enables the
determination of the magnitude as well as the sign of
\textbf{\textit{P}} with ARS, and the results are consistent with
those from the zero-field ARS. The experiments extend the utility of
field-split superconducting spectroscopy from tunnel junctions to
Andreev junctions of arbitrary barrier strengths.
\end{abstract}


\maketitle

\newpage

Superconducting spectroscopy has been one of the most effective
means of determining the spin polarization (\textbf{\textit{P}}) of
itinerant charge carriers in ferromagnetic materials. Two types of
electron transport in a superconductor (S)/ferromagnet (Fm) junction
can be used for this purpose: single particle tunneling \cite{a1}
and Andreev reflection (AR) \cite{a2}. AR \cite{a3}, which occurs at
an S/normal-metal (N) interface, is a process that converts the
quasiparticle current in N into supercurrent in S. In AR an incident
electron from the N side pairs up with an electron of opposite spin
and momentum to form a Cooper pair in order to enter the S, and a
hole is retro-reflected to conserve charge, spin, and momentum.
Therefore, AR results in a doubling of charge transfer across the
junction and an enhancement of the subgap junction conductance. In
an S/Fm junction, AR is suppressed due to the spin imbalance near
the Fermi level and the resulting reduction of the subgap Andreev
conductance can in principle be used to infer \textbf{\textit{P}}
\cite{a4}. In practice, in most cases both AR and normal reflections
are present and the zero-bias conductance alone does not give a
reliable measure of \textbf{\textit{P}}; one needs to measure and
analyze the entire conductance spectrum in order to separate the
effects of spin polarization and single electron tunneling, and
reliably determine \textbf{\textit{P}}. The analysis of the
conductance spectrum is done with a modified version of the
Blonder-Tinkham-Klapwijk (BTK) theory \cite{a5}, which takes account
of the spin polarization in the ferromagnet and computes the
junction conductance with a two-current (spin polarized and
unpolarized) model \cite{a6,a7,a8}. In the BTK theory the
probability of AR and normal reflection is determined by the barrier
strength, described by a dimensionless parameter $Z$, which includes
effects of physical scattering as well as band structure mismatches.
AR spectroscopy (ARS) has been widely implemented in point contact
setups \cite{a4,a8}, which has become an efficient technique for
rapid measurement of \textbf{\textit{P}} for a large variety of
ferromagnetic materials in various forms. However, there remain
several limitations and controversial issues with point contact ARS.
First, a point contact typically does not represent an interface in
a realistic device structure, while the magnitude and even the sign
of \textbf{\textit{P}} is known to depend on the nature of the
interface \cite{a9}. ARS, in general, only measures the magnitude of
\textbf{\textit{P}} and cannot determine its sign. Furthermore, the
fitting of the point contact ARS often requires an artificially
large spectral broadening \cite{a10} (or equivalently, the use of a
temperature in the Fermi function much greater than the actual
measurement temperature), and sometimes superconducting gaps much
different from the expected values \cite{a8,a10}. Finally, there are
ubiquitous observations of a precipitous decline of measured
\textbf{\textit{P}} with increasing $Z$ in a variety of systems
\cite{a8,a10,a11,a12}, which remain unexplained.

Single-particle tunneling in zero field cannot be used to measure
\textbf{\textit{P}} because of the degeneracy of the spin-up and
spin-down electrons. However, the application of an external
magnetic field lifts this degeneracy and the resulting asymmetry in
the conductance spectrum can be utilized to calculate the magnitude
and determine the sign of \textbf{\textit{P}} \cite{a1}.
Quantitative fits to the tunneling conductance spectrum with complex
structures are realized by using the coupled spin-up and spin-down
superconducting density of states (DOS) derived from the solution to
the Maki-Fulde equations \cite{a13}, which produces highly reliable
and unique \textbf{\textit{P}} values \cite{a14,a15}. Technically,
such spin-polarized tunneling (SPT) experiments are more challenging
to implement compared to ARS since they require fabrication of
high-quality tunnel junctions and a superconducting electrode with
high critical field and small spin-orbit coupling, which is in
practice limited to an ultrathin Al film.

In this paper, we report on the zero-field and Zeeman-split ARS
measurements of a series of doped-EuS/Al \emph{planar} junctions. By
controlling the growth temperature, the EuS films were naturally
doped to different levels due to varying degree of sulfur vacancies
\cite{a16}, which enabled realization of junctions of a relatively
wide range of intermediate $Z$ values where both AR and single
electron tunneling are prominent. We observe that the conductance
spectra can be fit straightforwardly (with \emph{zero} additional
spectral broadening and expected gap values) to the spin-polarized
BTK model. The fittings consistently yield \textit{\textbf{P}} of
$\sim$80 \% \emph{regardless of the $Z$ values}. Moreover, by using
planar junctions and thin Al counter-electrode, we are able to
obtain the ARS spectra in a large magnetic field. The Zeeman-split
ARS experiments have provided a means to extract the sign of
\textbf{\textit{P}} from ARS. It also demonstrates that the
field-splitting of the conductance spectra is not limited to tunnel
junctions but can be applied to S/Fm junctions of \emph{arbitrary
barrier strengths}, greatly simplifying its implementation. These
experiments have provided a reliable determination of the magnitude
($\sim$80 \%) and sign (+, majority spin polarized) of
\textbf{\textit{P}} for the doped EuS films.

EuS is a prototypical concentrated magnetic semiconductor. One of
the most attractive features found in such materials is a strong
exchange interaction between the spins of the itinerant charge
carriers in the conduction band and the localized magnetic moments.
This interaction is manifested as a giant spontaneous band splitting
of $\sim$ 0.5 eV \cite{a17}. Such materials offer high magnetization
and wide range of conductivity tunability so that they can be used
as spin filters \cite{a18,a19} in the insulating state and as spin
injectors when doped \cite{a20,a21,a22}.  Thus they offer an ideal
system to demonstrate the physics of semiconductor-based spintronic
devices in proof-of-concept studies.

\begin{figure}
\includegraphics[width=3.0in]{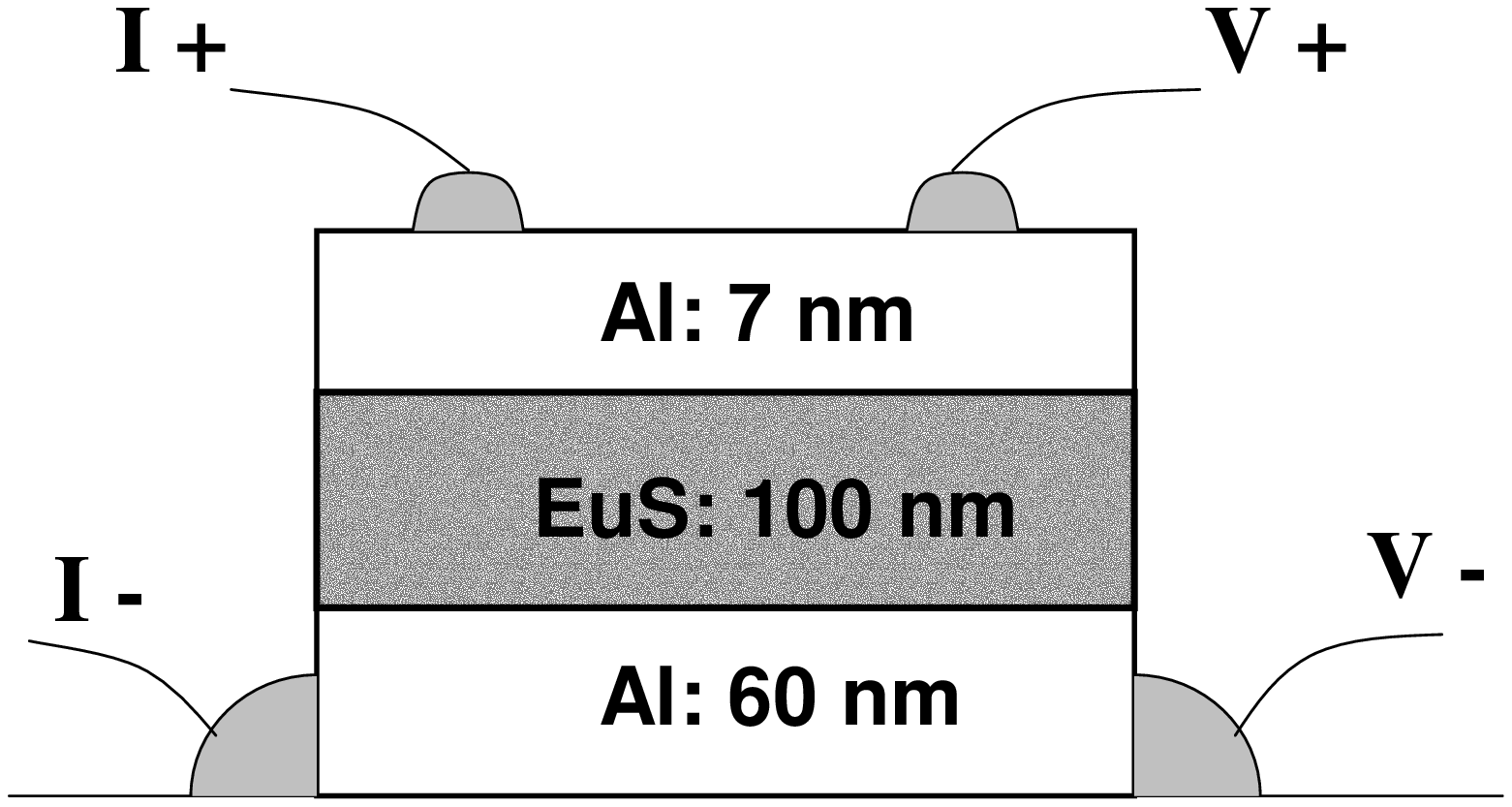}
\caption{A schematic diagram of the EuS/Al planar junction and the
contact scheme. The numbers are the typical thicknesses of the
films. The bottom junction of the thicker Al film and the EuS served
as a low-resistance Ohmic contact to minimize current crowding. }
\label{fig:fig1}
\end{figure}

\begin{figure}
\includegraphics[width=3.50in]{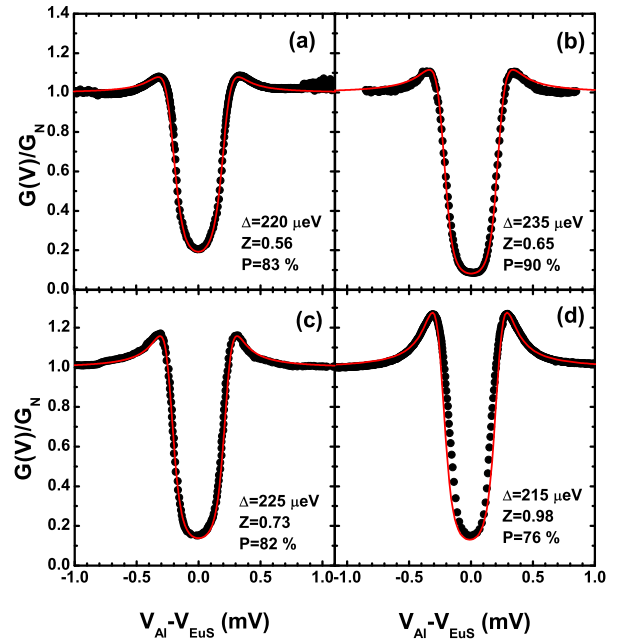}
\caption{Normalized conductance versus bias voltage of the
doped-EuS/Al junctions at a temperature of 0.38 K and zero magnetic
field. The growth temperatures for EuS films are: (a) -2 $^{0}$C;
(b) 34 $^{0}$C; (c) 80 $^{0}$C and (d) 120 $^{0}$C. The solid lines
are the best fits to the spin-polarized BTK theory. The fitting
parameters are indicated in the figures. }\label{fig:fig2}
\end{figure}

Doped-EuS/Al planar junctions were fabricated by vacuum deposition
on insulating Si(100) or glass (Corning) substrates. A schematic
diagram of the junction structure is shown in Figure 1. A relatively
thick (50-60 nm) Al stripe was first deposited. Conducting EuS films
of different conductivities, always 100 nm in thickness, were grown
at various low substrate temperatures by electron beam evaporation
in ultrahigh vacuum. The growth temperature was shown to be
effective in producing EuS films of varying doping levels, from
intrinsic to degenerate, by controlling the degree of sulfur
deficiency \cite{a16}. Finally, a thin Al electrode, 7-8 nm in
thickness, was thermally deposited immediately over the EuS as a
cross-stripe defined by a shadow mask. The effective junction
dimensions were $0.4 \times 0.4$ mm$^2$, and the junction
resistances at liquid Helium temperature varied from 3 to 15
k$\Omega$. The conductance spectra were obtained in a $^3$He system
using standard phase-sensitive lock-in detection. The EuS films used
in the present study had low-temperature resistivity on the order of
m$\Omega$ cm and carrier density of $\sim 10^{20}$ cm$^{-3}$; they
served as conducting electrodes rather than insulating tunnel
barriers.  The bottom Al/EuS junction made in this fashion always
resulted in a low-resistance Ohmic contact, which served to ensure
that there was negligible current crowding in the top junction. A
simple estimate of the resistance values shows that neither the EuS
film nor the bottom contact contributes significantly to the
measured resistance \cite{a23}. In addition, the application of a
small parallel field of about 1 kG, which fully suppresses
superconductivity in the thick bottom Al film but is much below the
critical field of the thin top Al electrode (at least 1.8 T), had
little effect on the conductance spectrum. This observation
demonstrated unambiguously that the measured conductance spectra
only reflected the top EuS/Al junction. The current ($I$)-voltage
($V$) characteristics of the junction at temperatures above $T_C$ of
Al shows a linear behavior. This is in stark contrast to EuS/In
junctions, which show a highly nonlinear $I-V$ characteristic of a
Schottky barrier \cite{a23}.

Shown in Figure 2 are the conductance spectra, d$I$/d$V$ as a
function of bias voltage $V$, for four EuS/Al junctions of different
barrier strengths in zero magnetic field. Each spectrum is
normalized by the corresponding one at a magnetic field above the
critical field for the Al. Qualitatively these spectra are
consistent with those of a S/Fm Andreev junction of intermediate $Z$
and large \textbf{\textit{P}} for the Fm, as judged from the much
diminished quasiparticle peaks near the superconducting energy gap,
$\pm \Delta$, and the low subgap conductance. These features are in
contrast to the case of pure tunneling in EuS/In junctions where a
Schottky barrier is present \cite{a23}. Quantitatively, these
spectra can be analyzed within the spin-polarized BTK model.
Excellent fits with physically sound parameters are obtained, as
shown in Figure 2. We emphasize that the fitting is always performed
in a straightforward manner and the only real adjustable fitting
parameters are $Z$ and \textbf{\textit{P}}. Firstly, $T$ in all of
the fits is always the actual measurement temperature; no additional
spectral broadening, either in the form an artificial $T$ higher
than the measurement temperature or an imaginary term in the
electron energy \cite{a24}, is necessary to obtain good fits. This
is evidence that Joule heating and inelastic effects including
magnetic pairing-breaking are immeasurably small in these junctions.
Secondly, the superconducting energy gaps for Al used are between
0.215 meV and 0.235 meV, values that are \emph{higher} than that for
bulk Al but expected of thin Al films \cite{a1}. The small variation
in the gap value is most likely due to differences in the Al
thickness. Thirdly, the \textbf{\textit{P}} values resulting from
these measurements and fittings show no substantial decline with
increasing $Z$, as shown in Figure 3 in which we plot
\textbf{\textit{P}} from five such junctions as a function of $Z$.
Within experimental uncertainty, there appears to be a small
decrease of \textbf{\textit{P}} with $Z$. However, this is in
contrast to the results from point contact ARS in many systems where
a much more significant decline of \textbf{\textit{P}} ($>50$ \%)
with $Z$ was observed in a similar $Z$ range \cite{a11}. We
attribute the small decrease in \textbf{\textit{P}} in our data to
actual changes of \textbf{\textit{P}} in films grown at increasing
substrate temperatures (from -2 $^{\circ}$C to 120 $^{\circ}$C),
which is known to reduce the EuS film conductivity \cite{a16}. This
result indicates that there is no intrinsic correlation between
increase of spin-flip scattering and $Z$ in these S/Fm junctions and
a natural transition to the limit of SPT is possible. We point out
that the above described observations, including the straightforward
excellent agreement with the spin-polarized BTK model and the
insensitivity of the determined \textbf{\textit{P}} with $Z$, are
not limited to the EuS junctions. Similar results have been observed
by us in junctions with the half metal CrO$_2$ \cite{a25} and the
ferromagnetic semimetal EuB$_6$ \cite{a26}.

In the BTK model, the parameter $Z$ includes physical (elastic)
scattering at the S/N interface as well as effects of band structure
mismatches. For example, the Fermi velocity mismatch results in an
effective barrier strength given by \cite{a27}
$Z_{eff}=\sqrt{(1-r)^2/4r}$, where $r$ is the ratio of the Fermi
velocities of the ferromagnet and superconductor. Under the present
growth conditions, it is estimated that the EuS has a carrier
(electron) density of $\sim 2.0\times 10^{20}$ cm$^{-3}$ at $T$=4.5
K \cite{a16}. Assuming a parabolic band and a unitary effective
electron mass, we estimate a Fermi velocity of $v_F^{EuS}=2.0\times
10^5$ m/s compared to $v_F^{Al}=1.8\times 10^6$ m/s for the
superconducting electrode Al. Such a large mismatch should result in
a substantial $Z_{eff}=1.35$ even in the absence of any physical
scattering at the interface. The small $Z$ values in our junctions
can be qualitatively attributed to enhanced junction transparency
due to a high spin polarization in the Fm electrode \cite{a28},
which has been widely observed in different S/Fm junctions of high
\textbf{\textit{P}} \cite{a29,a25,a30}. Another outstanding issue in
our data is the magnitude of the junction resistance, which is
several orders of magnitude higher than the prediction of the BTK
theory (for a ballistic point contact). The discrepancy has been
widely observed in S/semiconductor (Sm) junctions of different
materials and geometries \cite{a31,a32,a33}. Although a definitive
explanation of this observation is still lacking, it is expected
that the computation of the current and thus the junction resistance
should depend on the junction geometry and be different in planar
junctions \cite{a34}. It is important to note, however, that both in
our junctions and other S/Sm structures \cite{a31,a32,a33} the
conductance spectra are well described by the BTK theory.  This
represents a far more stringent requirement and strongly supports
its applicability in these structures. This assertion is further
reinforced by our results from measurements of the conductance
spectra under Zeeman-splitting magnetic fields.

\begin{figure}
\includegraphics[width=3.0in]{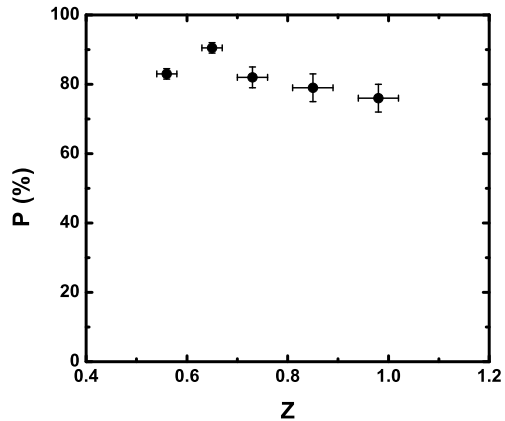}
\caption{The fitted \textbf{\textit{P}} as a function of $Z$ for
various doped-EuS/Al junctions with different EuS growth
temperatures.} \label{fig:fig3}
\end{figure}

\begin{figure}
\includegraphics[width=3.0in]{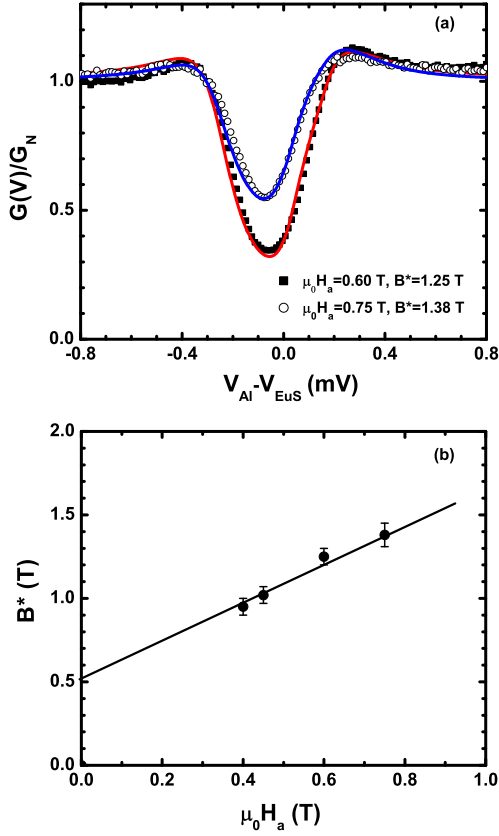}
\caption{(a) The Zeeman-split conductance spectra for a doped-EuS/Al
junction whose zero-field spectrum is shown in Figure 2(a). The
solid lines are the fits to the spin-polarized BTK theory that
incorporates the solution to the Maki-Fulde equations; (b) The
effective field B* used in the fittings as a function of the applied
field. The solid line is a linear fit to the data. }\label{fig:fig4}
\end{figure}

The use of a planar junction structure and thin Al electrodes afford
us the opportunity to Zeeman-split the superconducting DOS and
examine its consequences on the ARS spectrum. Figure 4(a) shows the
conductance curves of the EuS/Al junction of Figure 2(a) at in-plane
magnetic fields of 0.6 and 0.75 T.  Because the Al electrode was on
top of the EuS, it needed to be relatively thick (7-8 nm) which
resulted in a relatively low critical field ($<$2 T). However, even
these relatively low magnetic fields induce a sizable shift of the
conductance curve to the left. With the exception of noticeable
asymmetry near the peaks, there are no observable additional
features due to the minority spins. Qualitatively, these
observations indicate a large, \emph{positive}, \textbf{\textit{P}}
for the doped EuS. More importantly, to the best of our knowledge,
this experiment represents the first study of Zeeman-split ARS in
conventional S/Fm junctions.

In order to analyze the Zeeman-split ARS spectrum and independently
extract \textbf{\textit{P}} from the analysis, a thorough treatment
of spin-polarized charge transport in an Andreev junction with
Zeeman-splitting is necessary. This requires the use of the
appropriate spin-resolved DOS for Al in a magnetic field when
calculating the BTK transport (reflection and transmission)
coefficients (Table II in Ref. 5). The BTK coefficients depend only
on the parameter $Z$ and the coherence factors:
\begin{equation}
u_0^2=1-v_0^2=\frac{1}{2}\big[ 1+\frac{1}{N_S(E)}\big],
\end{equation}
where $N_S(E)$) is the normalized BCS DOS
\begin{equation}
N_S(E)=\frac{\mid E\mid}{\sqrt{E^2-\Delta^2}}.
\end{equation}
In a magnetic field $N_S(E)$ is Zeeman-split and the BTK
coefficients consequently become spin-dependent. Melin \cite{a35}
assumed a simple Zeeman-splitting of the BCS DOS in an applied field
and obtained the spin-dependent BTK coefficients using
\begin{equation}
u^2_{\uparrow(\downarrow)}=1-v^2_{\downarrow(\uparrow)}=\frac{1}{2}\big[1+\frac{\sqrt{(E\pm
\mu_BH)^2-\Delta^2}}{\mid E\pm \mu_BH\mid}\big],
\end{equation}
where $H$ is the applied magnetic field. The Zeeman-split
conductance curves at 0 K were then computed. This approach neglects
the effects of spin-orbit coupling and depairing from the applied
field. It has been shown \cite{a14,a15} that these effects are not
negligible even in a material such as Al. They result in significant
modification of the Zeeman-split conductance spectrum and,
particularly, ambiguity in the determination of \textbf{\textit{P}}
from it.  To obtain the DOS of a superconducting film in the
presence of spin-orbit coupling (parameter $b$) and pair-breaking
due to a magnetic field (parameter $\zeta$, which is proportional to
$H^2$), one needs to solve the Maki-Fulde equations:
\begin{equation}
u_{\pm}=\frac{E\mp \mu_BH}{\Delta}+\frac{\zeta
u_{\pm}}{\sqrt{1-u^2_{\pm}}}+b\big[\frac{u_{\mp}-u_{\pm}}{\sqrt{1-u^2_{\mp}}}\big].
\end{equation}
The solution of the coupled equations enables the determination of
the spin-resolved superconducting DOS,
\begin{equation}
\rho_{\uparrow(\downarrow)}=\frac{\rho(0)}{2}\textrm{Im}\big[\frac{u_{\pm}}{\sqrt{1-u^2_{\pm}}}\big],
\end{equation}
where $\rho_{\uparrow(\downarrow)}$ are the spin up(down)
superconducting DOS, $\rho(0)$ is the normal state DOS of the
superconductor at $E_F$. The spin-up (down) DOS can then be used to
calculate the corresponding spin-resolved BTK coefficients
\begin{equation}
u^2_{\uparrow(\downarrow)}=1-v^2_{\downarrow(\uparrow)}=\frac{1}{2}\big[1+\frac{1}{N_{S\uparrow(\downarrow)}(E)}\big],
\end{equation}
where
\begin{equation}
N_{S\uparrow(\downarrow)}=\textrm{Im}\big[\frac{u_{\pm}}{\sqrt{1-u^2_{\pm}}}\big].
\end{equation}
We numerically solve the Maki-Fulde equations (Eq. 4) and obtain the
actual DOS of the Al film in a magnetic field. The results are
similar to those obtained in Ref. 14 and used in the analysis of
Al/Fm tunnel junctions ($Z>>1$) \cite{a14,a15}. Using the DOS we
obtain the spin-dependent coherence factors (Eq. 6) and consequently
the BTK coefficients for different transport processes at arbitrary
barrier strength $Z$. We then calculate the junction conductance
under Zeeman splitting using these coefficients and the two-current
model \cite{Pedro}. This, therefore, is a general theoretical
framework that contains BTK \cite{a5}, spin-polarized BTK
\cite{a7,a8}, and Meservey-Tedrow \cite{a1} analysis as special
cases. It enables the quantitative analysis of the field-split
conductance spectrum of S/Fm junctions of \emph{arbitrary} barrier
strength.  As pointed out by Mazin \cite{Mazin}, ARS and
spin-polarized tunneling in general probe different forms of spin
polarization. In ARS, especially, depending on whether the electron
transport at the junction interface is ballistic or diffusive, the
spin densities are weighted differently by the Fermi velocities to
produce different current spin polarization. Our junctions are
clearly in the diffusive regime, and the measured
\textit{\textbf{P}} corresponds to a value with spin densities
weighted by $v^2_{F\uparrow(\downarrow)}$ (Eq. 2 of Ref. 37). In
Mazin's theory \cite{Mazin}, \textit{\textbf{P}} takes the same form
in the purely diffusive regime and when $Z >> 1$ (tunneling limit).
Thus a natural crossover exists between our case and the
Meservey-Tedrow regime \cite{a14,a15}.

The solid lines in Figure 4(a) are the best fits to the data using
the above scheme. The fits yield \textbf{\textit{P}} of 78\% and
73\% for applied fields ($\mu_0H_a$) of 0.6 T and 0.75 T,
respectively. In the fits the following parameters are used: $\zeta
=0.10$, $b$= 0.14 and effective magnetic field $B^{*}$ of 1.25 T and
1.38 T respectively. The parameter $Z$ (0.65) is determined
independently from the zero-field data [Figure 2(a)]. Although there
are a number of parameters in the fitting, the complexity of the
Zeeman-split conductance spectra makes the determination of the
parameters highly unique and reliable. The necessity to use an
effective magnetic field $B^{*}$ greater than the applied field is
readily apparent from the large shift of the conductance minimum
from the zero bias. In Figure 4(b) we plot $B^{*}$ as a function of
$\mu_0H_a$ (which are all greater than the saturation field of the
EuS). A linear fit of the data results in an intercept of 0.52 T at
$\mu_0H_a$= 0. These observations are consistent with the enhanced
Zeeman splitting in junctions where the Al films were in direct
contact with an insulating EuS barrier \cite{a18}. This enhanced
Zeeman splitting originates from the exchange interaction of EuS on
Al due to the intimate contact between them in these junctions. This
is to be contrasted with the case of tunnel junctions where the Al
is separated from the ferromagnet by a nonmagnetic insulator
\cite{a14}. This intimate contact also results in the large $\zeta$
and $b$ compared to those in pure Al, similar to the observation of
much enhanced spin-orbit interaction in thin Al with heavy
impurities such as rare earths \cite{a36} and noble metal [1] on the
surface. The \textbf{\textit{P}} determined from the fittings is
close to the value from zero-field ARS on the same junction, but
there appears to be a small but systematic decrease of the measured
\textbf{\textit{P}} with increasing magnetic field. This decrease in
\textbf{\textit{P}} is beyond the experimental uncertainty and
remains an open question.

In summary, we have performed a set of experiments to determine the
spin polarization of the magnetic semiconductor EuS using Andreev
reflection spectroscopy. Zero-field ARS on a series EuS/Al junctions
of different barrier strengths consistently yielded conductance
spectra that fit straightforwardly to the spin-polarized BTK model
and \textbf{\textit{P}} on the order of 80\% for the naturally doped
EuS, regardless of the barrier strength. Perhaps more importantly,
we have for the first time realized ARS in a large Zeeman-splitting
magnetic field in an S/Fm Andreev junction. The Zeeman-split ARS
spectra are well described via a modification of the BTK model to
incorporate the Al quasiparticle DOS in a magnetic field. The
zero-field results provide strong evidence for the applicability of
the spin-polarized BTK model to ARS in \emph{planar} S/Fm junctions
and the validity of its application for the determination of the
spin polarization of magnetic semiconductors. The experimental
realization of the Zeeman-split ARS and the development of a
theoretical framework for its understanding in junctions of
arbitrary barrier strength should greatly expand the utilization of
the field-split superconducting spectroscopy for the measurement of
the magnitude and sign of the spin polarization of ferromagnetic
metals and semiconductors.  The high \textbf{\textit{P}} in the
doped EuS films makes them an attractive source of spin-polarized
electrons in proof-of-concept spintronics studies.

The authors thank Prof. P. Schlottmann for his contributions in the
analysis of the Zeeman-split ARS and Prof. P. Stiles for helpful
discussions. One of the authors (C. Ren) would also like to thank
Ian Winger and Dan Read for technical assistance. This work was
supported by DARPA through ONR Grant Nos. N-00014-00-1094 and
MDA-072-02-1-0002.

*present address: National Laboratory for Superconductivity, Institute of Physics, Chinese Academy of Sciences, Beijing 100080

\# xiong@martech.fsu.edu

\end{document}